\documentclass[conference]{IEEEtran}
\IEEEoverridecommandlockouts
\usepackage{cite}
\usepackage{amsmath,amssymb,amsfonts}
\usepackage{algorithmic}
\usepackage{graphicx}
\usepackage{textcomp}
\usepackage{xcolor}
\usepackage{subcaption}

\def\BibTeX{{\rm B\kern-.05em{\sc i\kern-.025em b}\kern-.08em
    T\kern-.1667em\lower.7ex\hbox{E}\kern-.125emX}}

\DeclareMathOperator*{\argmax}{argmax}
\DeclareMathOperator*{\argmin}{argmin}
\DeclareMathOperator*{\softmax}{softmax}
\newcommand{\norm}[1]{\left\lVert#1\right\rVert}

\begin{document}

\title{Enhancing O-RAN Security: Evasion Attacks and Robust Defenses for Graph Reinforcement Learning-based Connection Management}

\author{\IEEEauthorblockN{Ravikumar Balakrishnan, Marius Arvinte, Nageen Himayat, Hosein Nikopour}
\IEEEauthorblockA{\textit{Security and Privacy Research} \\
\textit{Intel Labs}, USA\\
firstname.lastname@intel.com}
\and
\IEEEauthorblockN{Hassnaa Moustafa}
\IEEEauthorblockA{\textit{Network and Edge Group} \\
\textit{Intel Corporation}, USA\\
hassnaa.moustafa@intel.com}
}

\maketitle

\begin{abstract}
Adversarial machine learning, focused on studying various attacks and defenses on machine learning (ML) models, is rapidly gaining importance as ML is increasingly being adopted for optimizing wireless systems such as Open Radio Access Networks (O-RAN). A comprehensive modeling of the security threats and the demonstration of adversarial attacks and defenses on practical AI based O-RAN systems is still in its nascent stages. We begin by conducting threat modeling to pinpoint attack surfaces in O-RAN using an ML-based Connection management application (xApp) as an example. The xApp uses a Graph Neural Network trained using Deep Reinforcement Learning and achieves on average 54\% improvement in the coverage rate measured as the $5^{th}$ percentile user data rates. We then formulate and demonstrate evasion attacks that degrade the coverage rates by as much as $50\%$ through injecting bounded noise at different threat surfaces including the open wireless medium itself. Crucially, we also compare and contrast the effectiveness of such attacks on the ML-based xApp and a non-ML based heuristic. We finally develop and demonstrate robust training-based defenses against the challenging physical/jamming-based attacks and show a 15\% improvement in the coverage rates when compared to employing no defense over a range of noise budgets.
\end{abstract}

\begin{IEEEkeywords}
Adversarial Machine Learning, Wireless, O-RAN
\end{IEEEkeywords}

\section{Introduction}
Artificial intelligence (AI) and machine learning (ML) have been extensively studied for their applications to wireless communication systems and networks. Numerous bodies of work have recently demonstrated the benefits of applying AI/ML in the design, operation, and optimization of wireless systems. Early successes in applying modern deep learning to wireless systems include modulation classification~\cite{b1}, channel decoding~\cite{b2}, radio-frequency (RF) fingerprinting~\cite{b3} among many others. These techniques and the numerous ones that followed in the last half-decade have shown great promise and excitement stemming from the improved performance, automation, and re-usability of hardware. At the same time, there have been significant findings in mainstream signal domains where deep learning is applied (e.g., image, text, audio, etc.) demonstrating that deep learning-based models are vulnerable to adversarially crafted perturbations. This was first empirically demonstrated in~\cite{b4} where a bounded amount of human-imperceptible perturbation to the test samples leads a deep learning model to mis-classify these samples (e.g., mis-classifying an orange as an airplane). Adversarial machine learning has since garnered great attention and there have been numerous attack algorithms developed. There has also been progress in developing defenses to these attacks, with the most successful approaches leveraging adversarial training~\cite{b13, b14} -- a paradigm that re-formulates loss minimization as a robust min-max objective and introduces adversarial samples in the training loop. Other recent successful defensive approaches include randomized smoothing~\cite{b15} and diffusion-based pre-processing~\cite{b16}.

While there have been straightforward applications of adversarial machine learning to ML-based wireless systems such as in~\cite{b5, b6}, the nature of wireless environments also means there are unique threat actors that can inject noise via the environment itself. In particular, machine learning algorithms in wireless systems use raw or derived measurements of signals as input features. Such features are often received or measured by the ML algorithm over-the-air, which —while inherently involves benign receiver noise and stochastic variations due to the propagation characteristics -- are also additionally vulnerable to attackers that manifest their noise in the form of jamming signals over-the-air. In fact, this has been the focus of several recent approaches that study and tackle the impact of over-the-air attacks on radio-frequency based machine learning systems~\cite{b7, b8, b9, b10}. It is for this reason that the different adversarial threat models should be taken into consideration when designing the future generation of robust ML-based wireless systems.

\begin{figure*}[th]
    \centering
    \includegraphics[width=0.6\linewidth]{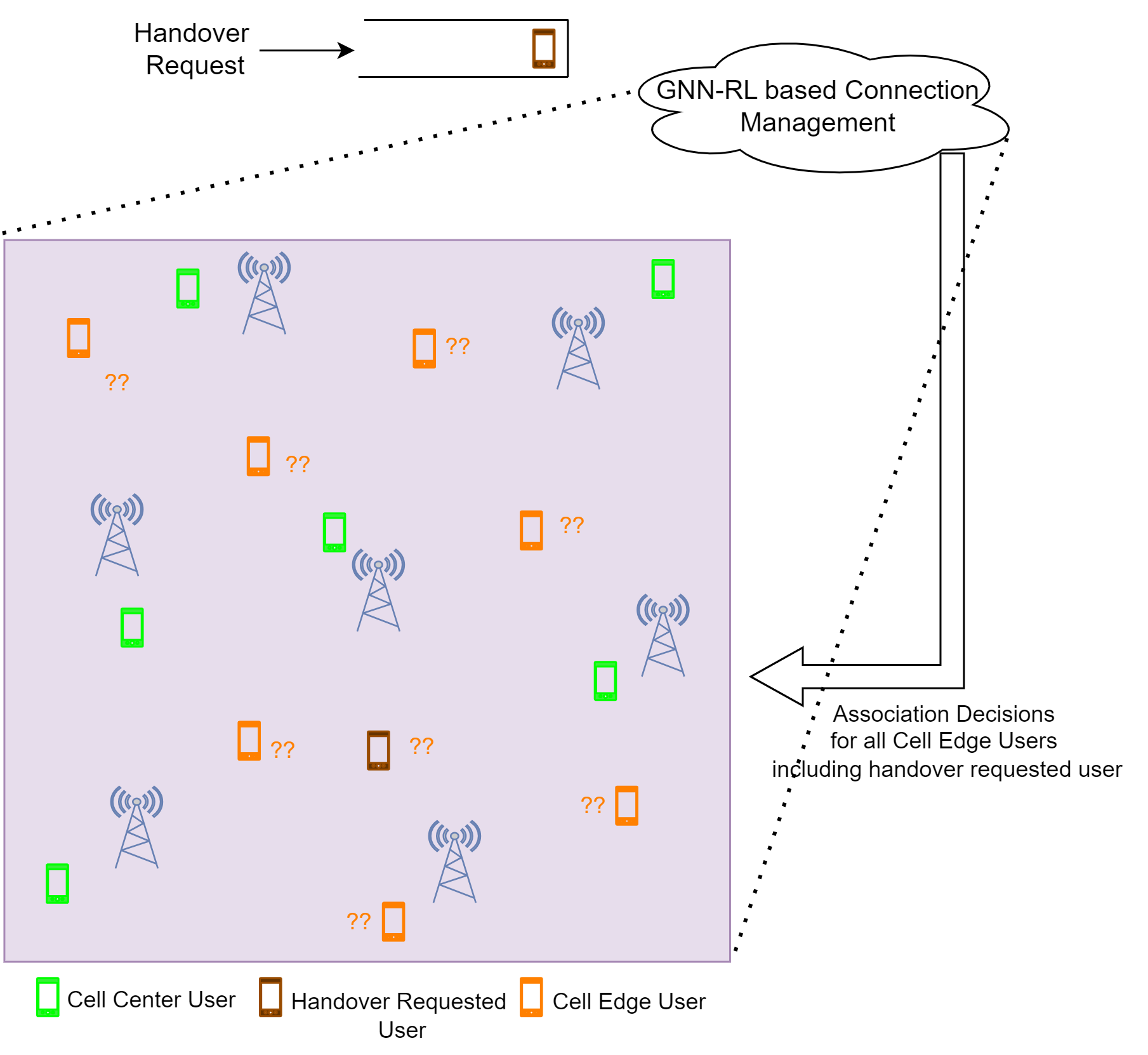}
    \caption{xApp utilizes GNN-RL to determine connectivity decisions for a handover requesting user and cell-edge users within a sub-graph around it.}
    \label{fig1}
\end{figure*} 

In this paper, we focus on the robustness of practical ML-based wireless systems, specifically pertaining to Open Radio Access Networks (O-RAN). O-RAN, in short, is the evolution of the traditional RAN architecture to an open, dis-aggregated, interoperable, virtualized, and AI-enabled architecture~\cite{b11}. One of the primary goals of O-RAN is to encourage a multi-vendor ecosystem that goes beyond the current paradigm of vendor ``lock-in" and allows the creation of new opportunities both for small and large businesses. A key aspect of the O-RAN architecture is its introduction of the RAN Intelligent Controller (RIC) that relies on AI techniques to embed intelligence into every layer of the O-RAN architecture. The RIC hosts AI/ML-based applications referred to as xApps and rApps to efficiently and automatically manage network resources. Examples of these applications include traffic steering, QoE prediction, and anomaly detection/correction, to name a few. Government agencies such as NSA and CISA have jointly released a document outlining the security considerations of O-RAN systems specifically raising awareness about AI related threats~\cite{b17}.

A graph neural network (GNN) based connection management xApp trained using reinforcement learning (RL) was proposed in~\cite{b12} and is the application targeted in this paper. Connection management comprises of decisions that involve keeping user equipment (UE) connected to the network during initial access, during mobility and handover events and also to achieve load balancing in the network. While these approaches have traditionally been handled in a UE-centric way (not globally), the O-RAN architecture allows the possibility of a global optimization and also the provision to utilize AI/ML techniques to that end at the RIC. As a result, in ~\cite{b12}, the authors utilize a graph neural network to abstract the O-RAN network as a graph where the cells and UEs form the nodes in an un-directed graph and the wireless link qualities between each user and each cell form the edge weights. The GNN computes a scalar quality metric indicating a score of any graph instance and is optimized using reinforcement learning. Reward signals that encode a score for the network drives the training and the resulting network has been shown to effectively achieve the objectives stated in the reward function as evidenced by extensive empirical findings. For the handover application, upon receiving a handover request, the xApp utilizes the GNN over a subnetwork around the handover requested user and obtains handover decisions for not only the requested user, but also cell-edge users in the sub-network as shown in Figure~\ref{fig1}.

An important issue remains, however, given that adversaries are able to inject noise in more than one ways into the features consumed by the ML model such that the resulting system performance may be seriously degraded. In this work, we comprehensively study this phenomenon of adversarial or evasion attacks and their corresponding defenses in the GNN-based connection management use-case. Specifically, we first identify three different threat models that allow for adversaries to inject noise into the system. Second, we develop new optimization objectives for adversaries and conduct a range of adversarial attacks under these threat models. Finally, we evaluate two defense techniques that can significantly improve the robustness/resilience of the ML model by modifying the training methodology. While connection management specifically is the target application considered in this paper, the attack and defense techniques demonstrated in this work can be extended to a range of wireless ML applications that operate on graph representation of data and also those that utilize reinforcement learning.

\section{System Model}
The system model consists of an O-RAN network with $N$ cells or base stations (BSs) and $M$ UEs, and a graph $\mathcal{G} = \mathcal{(V,E)}$ with $\mathcal{V}^{c} = \{v_0^c, \dots, v_{N-1}^c\}$ indicating the set of cells and $\mathcal{V}^{u} = \{v_0^u, \dots, v_{M-1}^u\}$ indicating the set of UEs. The edges between cells and UEs are denoted by $\mathcal{E}^{u} = \{e_{v_i^c, v_j^u} | v_i^c \in \mathcal{V}^{c}, v_j^u \in \mathcal{V}^{u} \}$. Further, the virtual edges between cells that convey information about their local networks are represented as $\mathcal{E}^{c} = \{e_{v_i^c, v_j^c} | v_i^c, v_j^c \in \mathcal{V}^{c} \}$. A virtual edge is assumed to be present upon satisfying a condition, for example, if the distance between the two cells is below a certain threshold. The set of UEs connected to a cell is given by $\mathcal{C}(v_i^c) = \{v_j^u |e_{v_i^c, v_j^u} \in \mathcal{E}^u, \forall j \}$.
The network topology is described at any time using two adjacency matrices $A_c \in \{0,1\}^{N \mathsf{x} N}$ and $A_u \in \{0,1\}^{N \mathsf{x} M}$ where
\begin{equation}
\mathbf{A}_{c} (i,j) =    \begin{cases}
1 & \text{if } e_{v_i^c,v_j^c} \in \mathcal{E}^c \\
0 & \text{otherwise}
\end{cases},
\label{adj1}
\end{equation}

\begin{equation}
\mathbf{A}_{u} (i,j) =    \begin{cases}
1 & \text{if } e_{v_i^c,v_j^e} \in \mathcal{E}^u \\
0 & \text{otherwise}
\end{cases}. 
\label{adj2}
\end{equation}

Given a configuration, the input features are constructed from the link quality measurement between cells and UEs. This is typically based on the reference signal received power (RSRP) measurements at UEs from different cells. A simple yet commonly adopted model, especially in power-limited scenarios such as mmWave communication links, is to approximate the link quality using the Shannon capacity equation as
\begin{equation}
c(v_i^c, v_j^u) = \log_2(1+\frac{P(v_i^c, v_j^u)}{N_0}) \textrm{[bits/sec]},
\label{capacity}
\end{equation}
\noindent where $P(v_i^c, v_j^u)$ denotes the RSRP at UE $v_j^u$ from cell $v_i^c$. The input features are constructed as functions of channel capacity matrix $\mathbf{C}$ and user rate matrix $\mathbf{R}$. The capacity matrix $\mathbf{C}$ is an $N\mathsf{x}M$ matrix containing elements $c(v_i^c, v_j^u)$. The rate matrix $\mathbf{R} \in \mathbb{R} ^{N \mathsf{x} M}$ measures the actual rate supported by the cell for each user based on the number of UEs connected to that cell and is given as $\frac{c(v_i^c, v_j^u)}{|\mathcal{C}(v_i^c)|}$. To effectively capture the relevant information for connectivity decisions, the following set of features are computed as functions of the cell-UE capacities, rates and the adjacency matrices~\cite{b12} and are given by
\begin{equation}
X_{c,1}^{(0)} = [\mathbf{A}_c \mathbf{R} \mathbf{1}_M || \mathbf{R} \mathbf{1}_M ] \in \mathbb{R}^{N\mathsf{x}2},
\label{feat1}
\end{equation}
\begin{equation}
X_{c,2}^{(0)} = [\mathbf{A}_u \mathbf{R}^T \mathbf{1}_N || \mathbf{C} \mathbf{1}_M ] \in \mathbb{R}^{N\mathsf{x}2},
\label{feat2}
\end{equation}
\begin{equation}
X_{u}^{(0)} = [\mathbf{C}^T \mathbf{1}_N || \mathbf{R}^T \mathbf{1}_N ] \in \mathbb{R}^{M\mathsf{x}2}.
\label{feat3}
\end{equation}
The vectors $\mathbf{1}_M$ and $\mathbf{1}_N$ represent all-one vectors of size $M$ and $N$ respectively and the $[\cdot || \cdot ]$ operator represents vector concatenation. The input features are fed to a graph neural network (GNN) which computes a $d$-dimensional latent feature for each vertex in the graph $\mathcal{G}$ at each layer. We utilize the same GNN architecture that was utilized in~\cite{b12}. Through $L$ layers of GNN processing, node features are propagated to other nodes such that each node's feature will contain aggregated information about its $L$-hop neighborhood. The feature vectors at the last layer of the GNN are combined to obtain a scalar score $Q(\mathcal{G})$ for the graph $\mathcal{G}$. 

\begin{figure*}[th]
    \centering
    \includegraphics[width=0.8\linewidth]{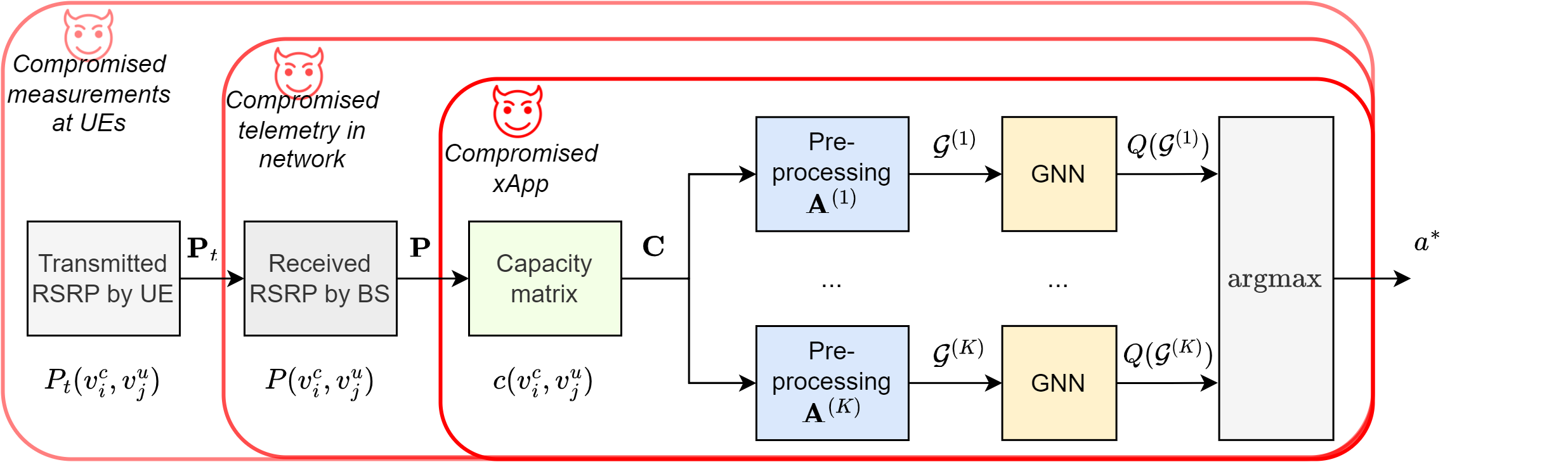}
    \caption{Information flow in connection management xApp highlighting threat surfaces to introduce adversarial perturbations. The three boxes indicate the three threat models.}
    \label{fig2}
\end{figure*} 

A deep Q-learning approach is utilized to learn the score function $Q(\mathcal{G})$, via learning the parameters of the GNN. This is achieved by observing samples of RAN deployments and the corresponding rewards obtained from the environment given set of actions. Training starts with a partially connected graph containing both connected and unconnected UEs. At each time step, the state $s_t$ is represented by the input features $X_{c,1}^{(0)}$, $X_{c,2}^{(0)}$, $X_u^{(0)}$ and the current adjacency matrices and a scalar score is obtained for each possible new connection management action $a_t = \mathcal{G}_{t-1} \cup e_{v_i^c, v_j^u}$ which represents connecting one un-connected UE from the current state $s_t$. One can therefore utilize $Q(\mathcal{G}_t)$ as a simplified notation of the Q value for taking action action $a_t$ from state $s_t$. A corresponding reward $r_t$ is calculated as the weighted sum of the improvement in network throughput $U(\mathcal{G})$ and the sum of minimum user rate at each cell, given by
\begin{equation}
r(s_t,a_t) = U(\mathcal{G}_t) - U(\mathcal{G}_{t-1}) + \frac{\lambda}{N} \sum_{i}^{N} \min_j c(v_i^c, v_j^u).
\label{reward}
\end{equation}
\noindent This allows us to balance between throughput improvement while encouraging a fair distribution of users across cells. Over the course of training, experiences in the form of $(s_t, a_t, r_t, s_{t+1})$ tuples are collected and stored in the replay buffer and utilized for updating the weights of the GNN using mean squared error as is commonly used in Deep Q Networks. When all UEs in the network are connected, the terminal state $s_T$ is reached.

In the xApp implementation, the above algorithm is implemented for handover application where mobile UEs request for new cell connections also referred to as handover events. To deal with the scale of cells and UEs, a local sub-graph of the O-RAN network is considered around a handover requested UE containing the cells from which the UE reported RSRP measurements and their L-hop neighbor cells. Following this, each UE in the sub-graph is classified as a cell-center UE or a cell-edge UE based on a certain RSRP threshold. Thus, the initial connectivity graph $\mathcal{G}_0$ contains a set of cell-edge UEs also termed as reshuffled UEs that require new cell connections.

\section{Threat Modeling and Adversarial Attacks on Connection Management xApp}
Evasion attack is one of the most common attacks on the ML model where an adversary injects a small amount of noise to the model's input features such that it is easy to evade detection while causing a large deviation to the model output (usually accompanied with degradation in performance).

In this work, we aim to test the robustness of the connection management xApp to such evasion attacks and then develop the corresponding defenses. In order to obtain a deeper understanding, we first study the information flow in the xApp pipeline to see how an adversary can introduce perturbations to the ML model's input features. Figure~\ref{fig2} details this in the context of O-RAN system. At the core of it, the UEs measure their received powers over the resource elements on the reference signals transmitted by their cells or RUs. Each UE measures the RSRP from a set of cells and share this information over a control channel to their serving cell. The cells, in turn, aggregate this information and send it to the RIC that hosts the xApp. The xApp further computes the capacity matrix C, rate matrix R and the binary adjacency matrices in order to compose the input features for the GNN models. Since at any time $t$, all possible/valid actions are evaluated, we have $K_t$ GNN instances that are executed followed by an $\argmax (\cdot)$ function selecting the action that maximizes the network score.

\subsection{Threat Models}
Figure~\ref{fig2} illustrates three distinct threat models that can give rise to perturbed inputs flowing to the GNN models.

\begin{enumerate}
	\item \textbf{Compromised xApp:} When the xApp is compromised or acting maliciously, an adversary can manipulate inputs $X_{c,1}^{(0)}$, $X_{c,2}^{(0)}$ and $X_u^{(0)}$ by introducing bounded adversarially-crafted additive noise matrices $\Delta X_{c,1}^{(0)}$, $\Delta X_{c,2}^{(0)}$ and $\Delta X_u^{(0)}$.
	\item \textbf{Compromised Network Telemetry:} Another entry point for an attacker is to conduct man-in-the-middle attacks and introduce the additive noise to the RSRP telemetry between cells $\Leftrightarrow$ RIC, RIC $\Leftrightarrow$ xApp or xApp $\Leftrightarrow$ xApp interfaces and APIs.
	\item \textbf{Compromised measurements at UEs:} In this unique attack on wireless systems, adversaries can introduce bounded additive noise $\Delta P(i,j)$ to the RSRP measurement $P(i,j)$ on one or more cell-UE links. We further categorize this into a case where adversaries can perturb all UEs' RSRP measurements, and a case where they perturb only a subset of UEs' RSRPs.
\end{enumerate}

\subsection{Adversarial Attacks on Connection Management xApp}
The goal of the attacker is to disrupt the network score function $Q(\mathcal{G})$ such that the xApp causes degradation in the network performance. In order to realize this, we use the following design considerations while developing the attacks.

\subsubsection{Attacker Objective}
Attacks can be classified as white-box and black-box attacks depending on whether the attacker has access to the model and its parameters. The former case (white-box) is a stronger form of attack and also difficult to defend against. In order to design a white-box attack, one needs to define an objective function for the attacker to optimize. Black-box attacks can be of many types one of which being the case where the attacker uses a surrogate model with only access to the model architecture but while being able to utilize gradient-based attacks. 

In either of the above cases, one can define an objective function that maximizes the error in the network score function $Q(\mathcal{G})$. However, such an approach has two challenges. Firstly, since the xApp utilizes a reinforcement learning approach, we do not have the ground truth for the network score function. Secondly, disrupting the score function may impact the network score for the optimal action but may also assign a high score for any other sub-optimal, yet good action. For instance, connecting a UE to its second-best cell may be a valid outcome of such an attack yet may not cause significant disruption to the network.

\begin{figure*}[t]
    \centering
    \includegraphics[width=0.9\linewidth]{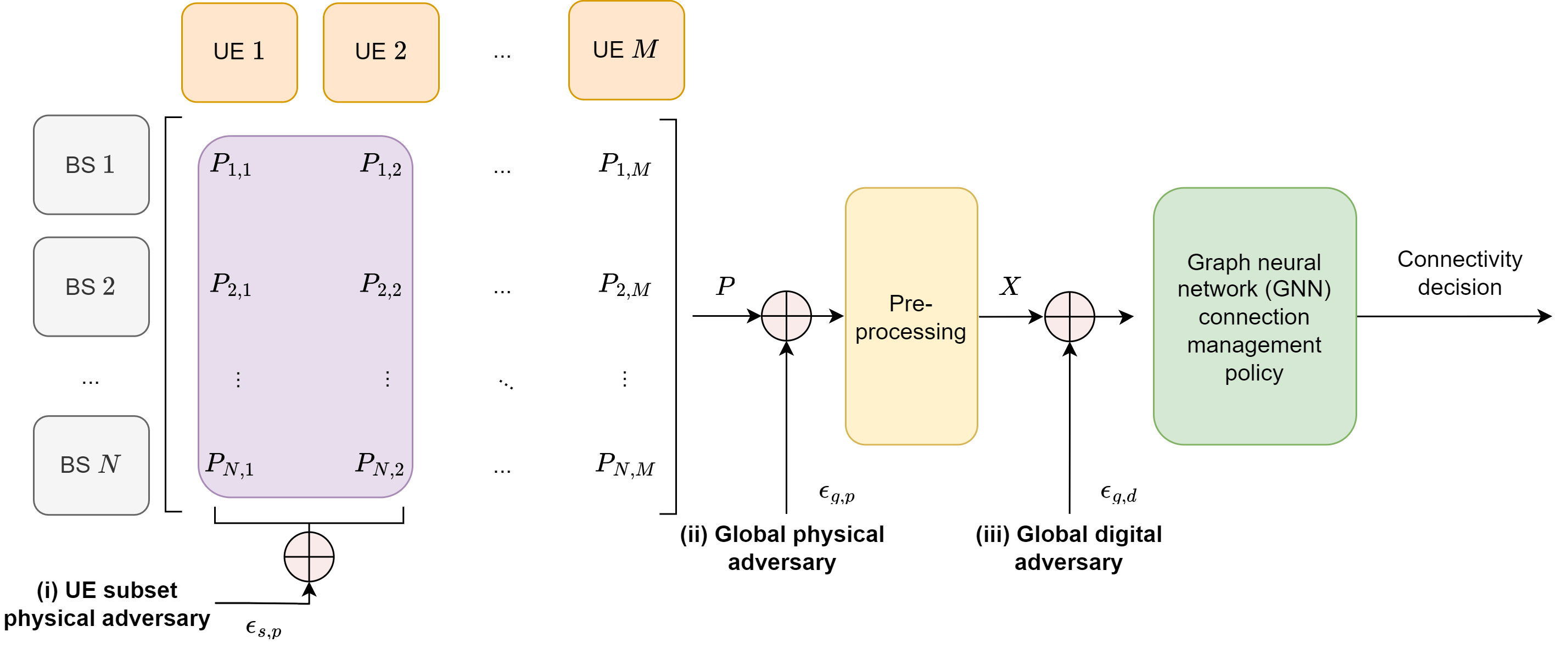}
    \caption{Illustration of the three attacks investigated in this paper, shown from left to right, in increasing order of their capabilities and threat model strength: (i) a physical adversary that can only perturb the RSRP measurements for a subset of UEs, (ii) a physical adversary that can perturb the RSRP measurements for all UEs, and (iii) a fully digital adversary with direct access to the GNN inputs.}
    \label{fig3}
\end{figure*}

To address the issue of a lack of ground truth, an attacker may construct logits from the outputs of the $K$ copies of GNNs and then apply $\argmax(\cdot)$ to obtain the labels. To overcome the concern about executing an effective attack, in this work, we construct a targeted attack which aims to overload a few cells such that per-UE data rates are severely impacted while leaving other cells to be underutilized. This is achieved by maximizing the log likelihood of the probability of selecting the same cell for every UE in the network (whenever possible). To tackle this, we first link the $K$ GNNs' by applying a $\softmax(\cdot)$ function to their outputs to produce probabilities over the $K$ actions. The $\softmax(\cdot)$ function essentially approximates $\argmax(\cdot)$ of the victim's processing chain shown in Figure~\ref{fig2} while providing both the ground truth labels as well as allowing differentiability through the input processing chain. We do not require the ground truth as by simply forcing as many UEs to the fewest cells as possible, we can achieve the desired degradation in terms of the coverage rate. Therefore, the targeted attack can be formulated by the adversary by maximizing the objective function
\begin{equation}
\label{att_general}	
\begin{split}
s_t^*  & = \argmax_{s_t} \log \left( \hat{Q}^\mathcal{(T)}(\mathcal{G}_t^{(1)}, \dots, \mathcal{G}_t^{(K)}) \right), \\
& \quad \quad \textrm{s.t.} \quad \norm{s_t^* - s_t}_p  \leq \epsilon
\end{split}
\end{equation}
\noindent where the optimal perturbed state $s_t^*$ is obtained by maximizing the log probability corresponding to the $\softmax$ output $\hat{Q}^\mathcal{(T)} = \frac{\exp(Q (\mathcal{G}_t^{\mathcal{(T)}})}{\sum_{k=1}^K \exp(Q(\mathcal{G}_t^{(k)}))}$ of a target action $\mathcal{T}$ subject to an $L_p$ norm bound on the perturbation $\epsilon$. The above framework allows for a range of target actions depending on the attacker's goals. The target action we employ in our approach are those actions leading UEs to a given macro-cell (among a set of macro-cells and small cells). One can easily use the above formulation to define other goals such as selecting the worst cell-UE pair. To achieve this, one may simply substitute the target action as the one that has the lowest Q-value probability $\argmin_a (\hat{Q}(s_t, a_t))$. It also allows use of standard solvers that rely on gradient-based techniques such as projected gradient descent (PGD). We note that the number of possible actions $K$ varies with time and hence the number of copies of GNNs for the attacker at any given time. We investigate three different attacks which we summarize in Figure~\ref{fig3}. The major differences among them arise from the attacker's capability in terms of whether they can directly attack the digital inputs to the models or whether they can only perform jamming attacks on either all or a subset of UEs' RSRP measurements. These are described in greater details as follows.

\subsubsection{Digital Attacks}
The most common attack that has been studied in the field of adversarial machine learning is a digital attack directly on the inputs to an ML model with an upper bound on the max norm of the noise vector. In computer vision, this limits the extent to which each pixel value is modified and therefore allowing the noise to be imperceptible to the human eye. When the xApp is compromised, attacker is able to directly modify the inputs to the GNN. The inputs to the GNN basically contains a set of matrices defined in equations~(\ref{feat1})-(\ref{feat3}) for each of the $K$ GNNs where the entries contain measures of sum rates across neighboring cells and UEs. For notation simplicity, we use the $K$ graphs to represent the perturbed sets of matrices, with the attack given by the constrained optimization
\begin{equation}
\label{dig_attack}
\begin{split}
 \{\mathcal{G}_t^{(1)*},\dots,& \mathcal{G}_t^{(K)*}\} \\ = &\argmax_{\mathcal{G}_t^{(1)}, \dots, \mathcal{G}_t^{(K)}} \log \left( \hat{Q}^\mathcal{(T)}(\mathcal{G}_t^{(1)}, \dots, \mathcal{G}_t^{(K)})\right),
\end{split}
\end{equation}
\noindent Utilizing a max norm allows the noised aggregate capacities and rates to be bounded by a threshold.

\subsubsection{Physical/Jamming Attack}
The open nature of wireless medium means that adversaries can inject noise with the help of inexpensive radio equipment thus impacting a receiver that is running the ML model. More concretely, in a typical cellular system, a base station transmits reference signals (e.g., demodulation reference signals) so that UEs can perform channel estimation, channel quality prediction and report it back to the base stations. This information will be utilized for a range of tasks such as resource allocation, multi-antenna communications, connection management among others. Since the time-frequency positions of these references signals are typically standardized, an adversary can transmit jamming signals such that the channel quality predictions are distorted. In a simple scenario, measurements of received power on the reference signals (RSRP) by the UEs from the base stations can be distorted by adversarial noise which then is propagated to the base station or RIC that is running the ML algorithm. This is illustrated in Figure~\ref{fig3}.

It is important to note that unlike the usual jamming attacks that may require a significant noise power to disrupt a receiver, ML based systems have been shown to require a small amount of perturbation (usually bounded by L2 norm or max norm) to achieve large degradation. To this end, we first define a quantity Perturbation-to-Noise ratio (PNR) as the ratio of the perturbation power to the receiver noise power and is expressed in dB units in log scale. An attacker aims to achieve perturbation bounded by the PNR value at UEs’ receivers such that the RSRP measurement (in dBm) is disrupted by at most PNR units.

For an attacker conducting physical attack on the CM xApp, the attacker will perturb the RSRP measurement by at most PNR units. The attacker's objective, concretely, is then defined as the solution to the constrained optimization objective
\begin{equation}
P_t^* = \argmax_{P_t} \log \left(\hat{Q}^\mathcal{(T)}(\mathcal{G}_t^{(1)}, \dots, \mathcal{G}_t^{(K)}) \right),
\label{eq10}
\end{equation}
\noindent where the perturbation to individual entries of the RSRP matrix are constrained by $\epsilon_p$ dB which indicates the PNR. Depending on the capability of the attacker, the attacker may impose further constraints such as the number of UEs and/or cell-specific RSRP that the attacker can perturb. The same RSRP matrix $P$ is passed through to the xApp and goes through pre-processing before being passed to each of the $K$ GNN models. Since the entire processing chain shown in Figure~\ref{fig2} is differentiable (after replacing $\argmax(\cdot)$ with the $\softmax(\cdot)$ function), an attacker can replicate the processing chain and compute the optimum perturbation using gradient-based approach such as PGD.

While a physical attack can be effective to disrupt the network performance, in practical settings, it becomes increasingly difficult for an attacker to execute attacks across all the UEs in a network. In fact, it is very likely that the attacker may transmit jamming signals over only a set of UEs (typically in the same region). To evaluate such an attack, we construct a patch attack where a binary mask is placed over the RSRP matrix P such that the adversary can only impact the masked set of UEs.

\section{Defending against Adversarial Attacks on Connection Management xApp}
\label{defenses}

\subsection{Hardware based Defense}
Confidential Computing can greatly benefit from adversaries accessing model inference at runtime by running the xApp code within a secure enclave such as Trusted Execution Environments. As a result, an adversary does not have access to the model information such as the model architecture, trained parameters, etc. This forces the adversary into black-box operation which has traditionally shown to be less effective than a white-box attack. Furthermore, by hardening the xApp along with encrypting telemetry information between the different network entities, one can ensure that the raw data is only accessed within the secure enclave by the xApp while running the ML model.

\subsection{Algorithmic Defenses}
\begin{figure}[thb]
    \centering
    \includegraphics[width=0.9\linewidth]{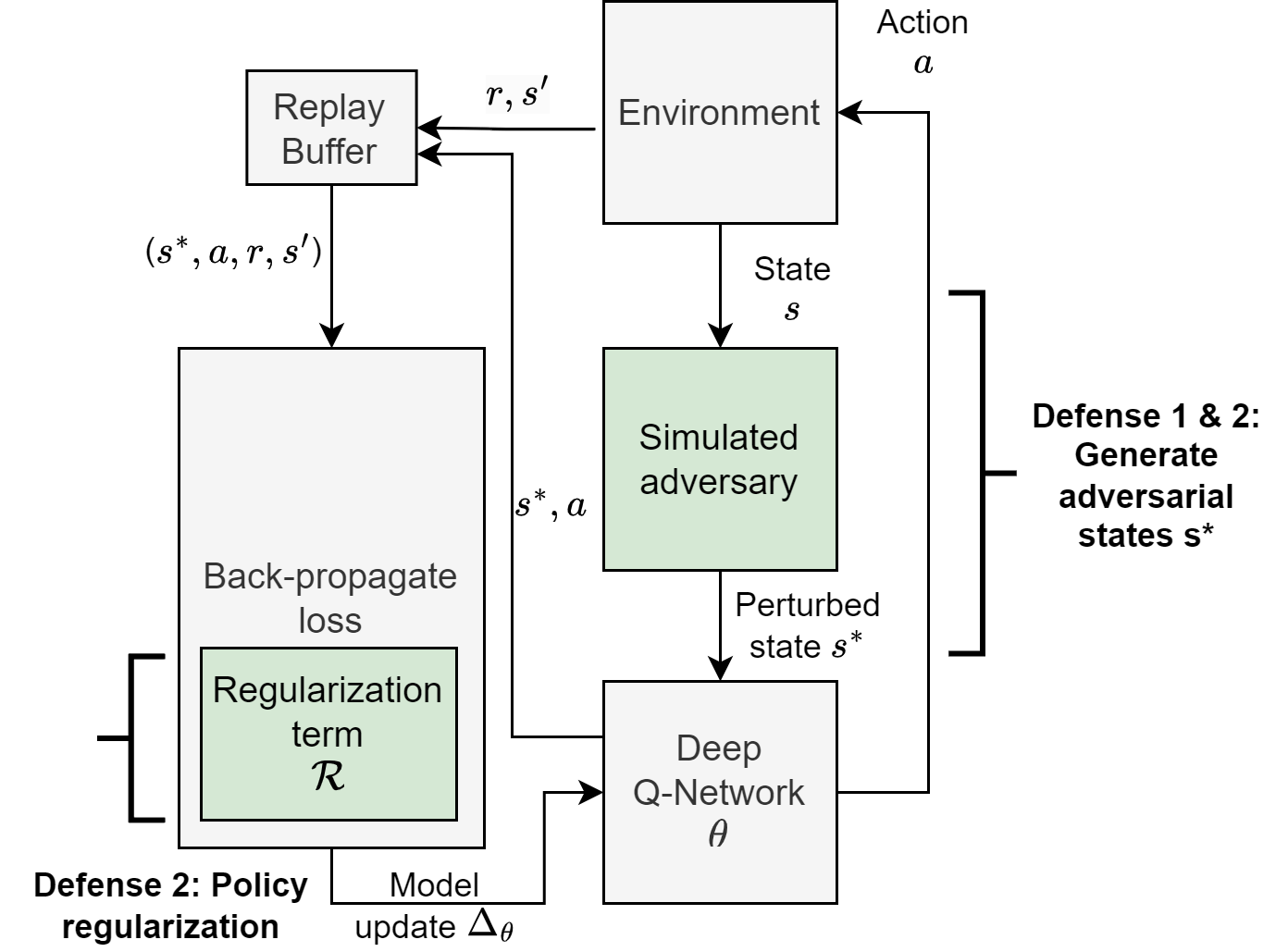}
    \caption{Illustration of the two proposed defenses for the RL-based allocation policy.}
    \label{fig:defense_diagram}
\end{figure}

While confidential computing can be effective against attacks such as the digital attack, it cannot stop an adversary from perpetrating physical attacks by injecting noise into the wireless medium during the RSRP measurement phase. As a result, it is imperative to consider strong algorithmic defenses that can inherently build robust/resilient ML models. We evaluate 2 algorithmic defense approaches at training time in order to overcome the coverage degradation caused by physical attacks as illustrated in Figure~\ref{fig:defense_diagram}.

\subsubsection{Adversarial Training} 
In this method, we augment the training set with samples generated by conducting adversarial attacks using PGD algorithm. Given the sequential and episodic nature of reinforcement learning, we apply the adversarial training in a way that is practical for RL settings. We first roll out trajectories under benign setting to generate state-action-reward tuples to be stored in the replay buffer for the RL agent and train using the DQN approach. Then subsequently, we conduct PGD attacks on the states and follow the trajectory of perturbed state transitions to collect a new set of samples to further fine-tune the model using the adversarial samples. We varied the PNR bound as a hyper-parameter for training in order to ensure robustness of the model over a range of validation scenarios and noise levels. This method has the advantage that one can take a pre-trained model and only fine-tune it with the help of perturbed experience tuples.

\subsubsection{Regularized Training}
An important observation in our RL setup is that although the adversary is able to modify the underlying state (RSRP measurement) for the xApp, the underlying data channels are not impacted by the adversarial noise. Therefore, while the state does get modified from the xApp perspective, the underlying RL environment’s state transitions are governed by the original state but with the new sub-optimal actions. In this light, we utilize recent efforts in developing theoretically-principled defense via regularized training. In this approach, the adversaries are assumed to make the state ``appear" perturbed instead of actually modifying it~\cite{b18}. The authors in~\cite{b18} also show that the regularized training approach inherently aims to bound the distance between the optimal policy under the original unperturbed states and the sub-optimal policy that uses the perturbed states. Since for the case of DQN, the policy is simply an $\argmax(\cdot)$ function, we utilize a hinge-like policy regularizer defined as
\begin{multline}
\mathcal{R}_{DQN}(\theta) := \sum_P \max  \{ \\
\max_{P^* \in B(P)} \max_{a \notin a^*} \bar Q_{\theta}(P^*, a) - \bar Q_{\theta}(P^*, a^*(P)), -c \}.
\label{eq11}	
\end{multline}
\noindent We compute this term by dynamically constructing $K$ GNNs, obtaining the worst-case perturbation using the PGD algorithm and then calculating the distance between the likelihoods between the benign and worst-case setting. The perturbation bound $B(P)$ is captured by the PNR bound that only allows the RSRP to be modified at most by PNR dB. We then utilize the regularizer along with the original DQN objective function to update the parameters of the DQN. The trained model then replaces the original model as a robust version with guarantees on the performance loss in a way that bounded perturbations can lead to bounded divergence between the policies with and without perturbation. This is due to the fact that the regularizer increases the likelihood of the top-1 actions to be identical with and without bounded perturbations.

\begin{figure}[ht]
    \centering
    \includegraphics[width=0.9\linewidth]{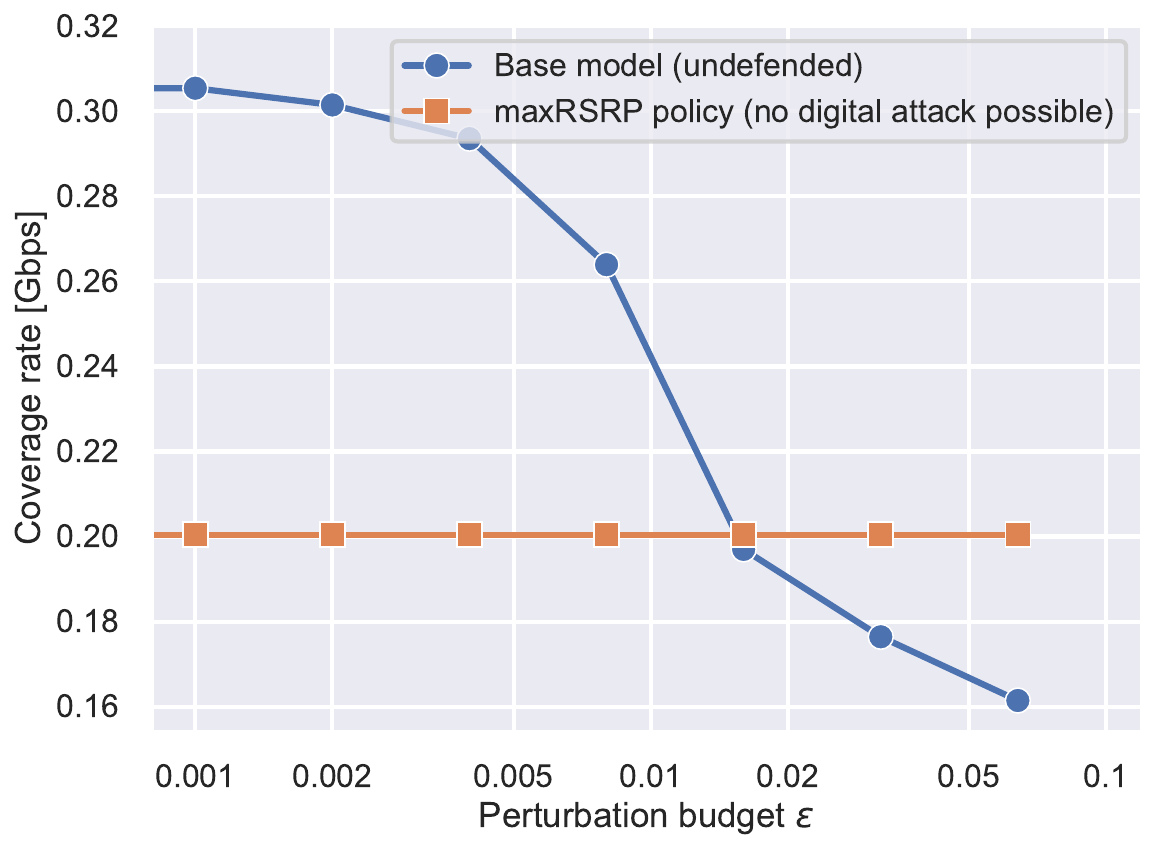}
    \caption{Illustration of the impact of digital attack on the coverage performance. Adversary is assumed to add noise directly to the input feature of the model.}
    \label{fig4}
\end{figure}

\section{Experimental Evaluation}

\begin{figure*}[t!]
    \centering
    \begin{subfigure}[t]{0.47\textwidth}
        \centering
        \includegraphics[width=0.8\linewidth]{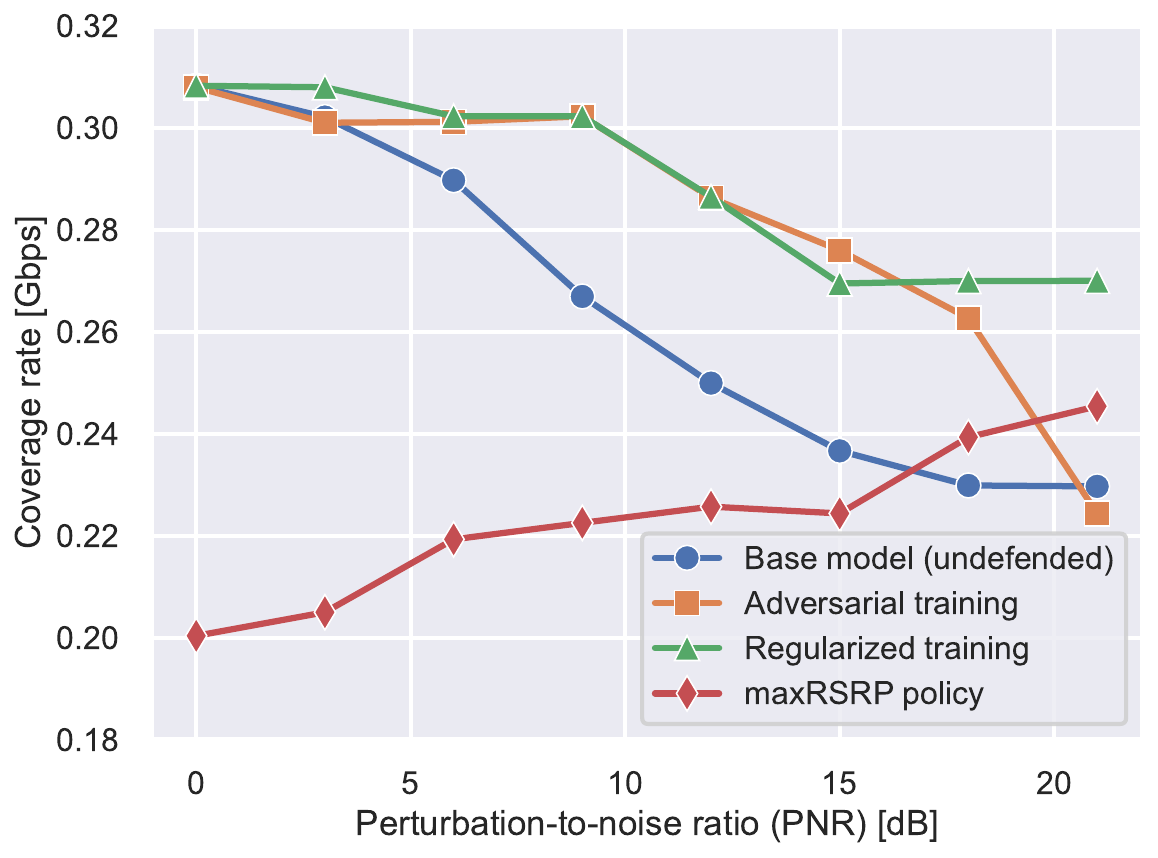}
        \caption{Whitebox PGD attack on RSRP measurements.}
        \label{fig5a}
    \end{subfigure}
    \begin{subfigure}[t]{0.47\textwidth}
        \centering
        \includegraphics[width=0.8\linewidth]{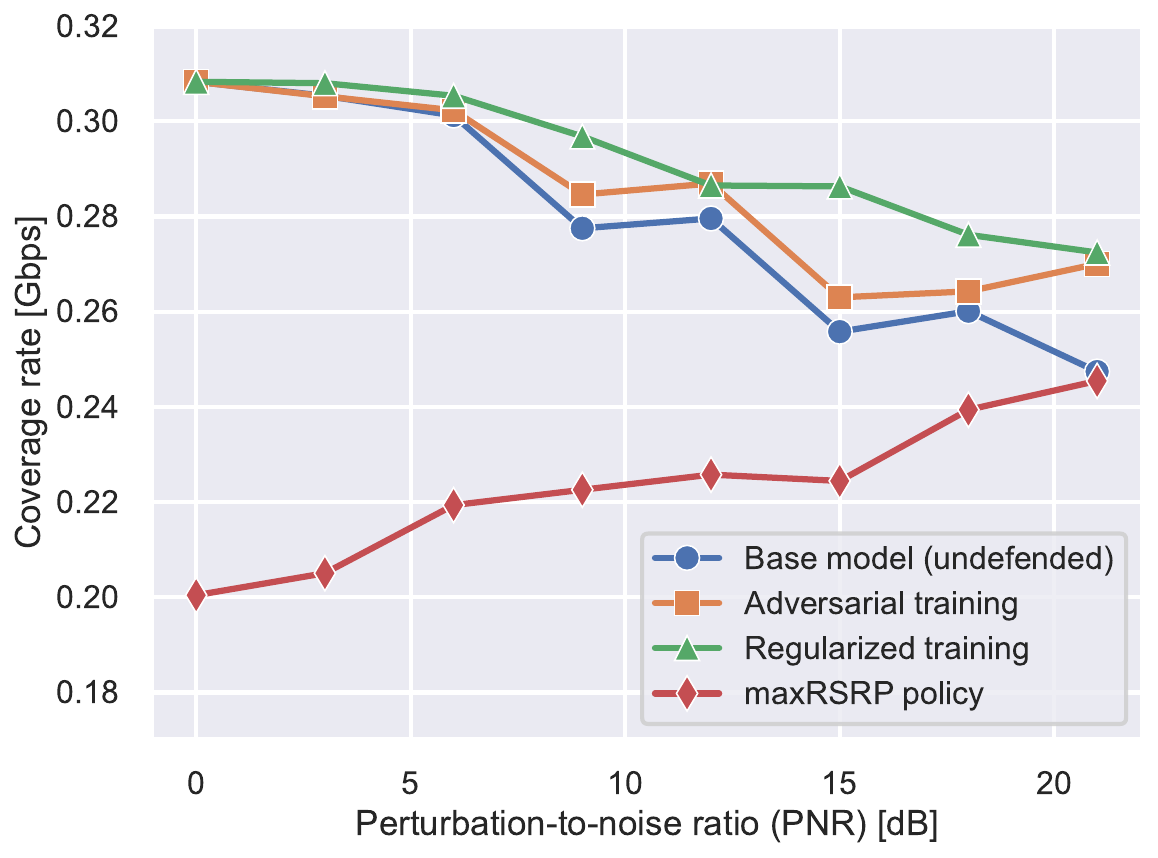}
        \caption{Blackbox attack using attacker's surrogate model}
        \label{fig5b}
    
    \end{subfigure}
\caption{Coverage rate (Gbps) as a function of PNR (dB) using Physical Attacks (whitebox and blackbox)}
\end{figure*}

\subsection{Attack Evaluation}
We evaluated the performance of the xApp under both digital and physical attacks using a simulated network of cells and UEs. We created five different scenarios each of which contains a unique deployment of $N=6$ cells and $M=50$ UEs. It is assumed that all UEs have RSRP measurements from the neighboring cells and for the missing ones, we set them to $-\infty$. Each UE was classified as a cell edge UE or a cell center UE using a threshold criteria as defined in~\cite{b12}. All cell edge UEs were assumed to require handover/cell reassignment decisions. Thus the initial graph $\mathcal{G}_0$ contains the connections between cells and cell center UEs while the connections between cells and cell edge UEs are sequentially determined using the connection management xApp as shown in Figure~\ref{fig2}. The GNN architecture utilized is the same as defined in~\cite{b12} with $L=2$ layers and $d=8$ dimensions per layer.

We utilized PGD for conducting gradient-based attacks. The adversary replaces the original model inputs using the noised version using PGD and passes the perturbed input to the model to execute. In the case of digital attacks, the noise added to each feature is upper bounded by max norm bound of $\epsilon_d$ where the features are normalized between 0 and 1. In the case of physical attacks, the adversary perturbs the RSRP matrix $P$ using max norm bound on the PNR resulting in a perturbed matrix $P^*$.

Figure~\ref{fig4} shows the impact of digital attack which is also the strongest attack by plotting coverage rate as a function of the noise budget epsilon. For reference, the plot also shows the coverage rate of a maxRSRP algorithm which greedily selects the cell with the highest RSRP for each UE. In benign settings with no adversarial noise, the GNN offered over $54\%$ improvement over the maxRSRP algorithm. However, as epsilon budget increased, the coverage rate of GNN based approach steadily degraded reaching a minimum of $0.16$Gbps well below the maxRSRP algorithm’s $0.2$ Gbps. The minimum value of $0.16$Gbps (~48\% drop in coverage rate) is also reached with a relatively low epsilon budget of $0.064/1$ which is about $6\%$ of the max value per feature.

In Figures~\ref{fig5a}, ~\ref{fig5b}, we show the impact of physical attacks where the adversary directly perturbs the RSRP matrix both under a white-box attack and a black-box attack using a surrogate model. To allow for a greater challenge to the adversary, we initialize the adversary's surrogate model with random weights. We varied the PNR in units of $3$dB as this is roughly equivalent to doubling the noise power in Watts at the receiver. For reference, received power is typically few tens of dB above the noise floor. By perturbing the reference signals, the attacks effectively raises the noise floor by PNR dB, thus impacting the capacity estimate obtained in Equation~\ref{capacity}.

The degradation in coverage rate with increase in PNR budget is evident for the undefended base model in the case of white-box attack where low PNR budget such as 9dB attack can lead to $\sim13\%$ drop in coverage rate and even stronger attacks are sufficient to almost erase all the gains of the Connection management approach over the maxRSRP policy. The black-box attack also leads to  $\sim10\%$ drop in coverage rate for a 9dB attack, but does not scale sufficiently to eliminate the gains of the ML policy.

\begin{figure}[htbp]
    \centering
    \includegraphics[width=0.9\linewidth]{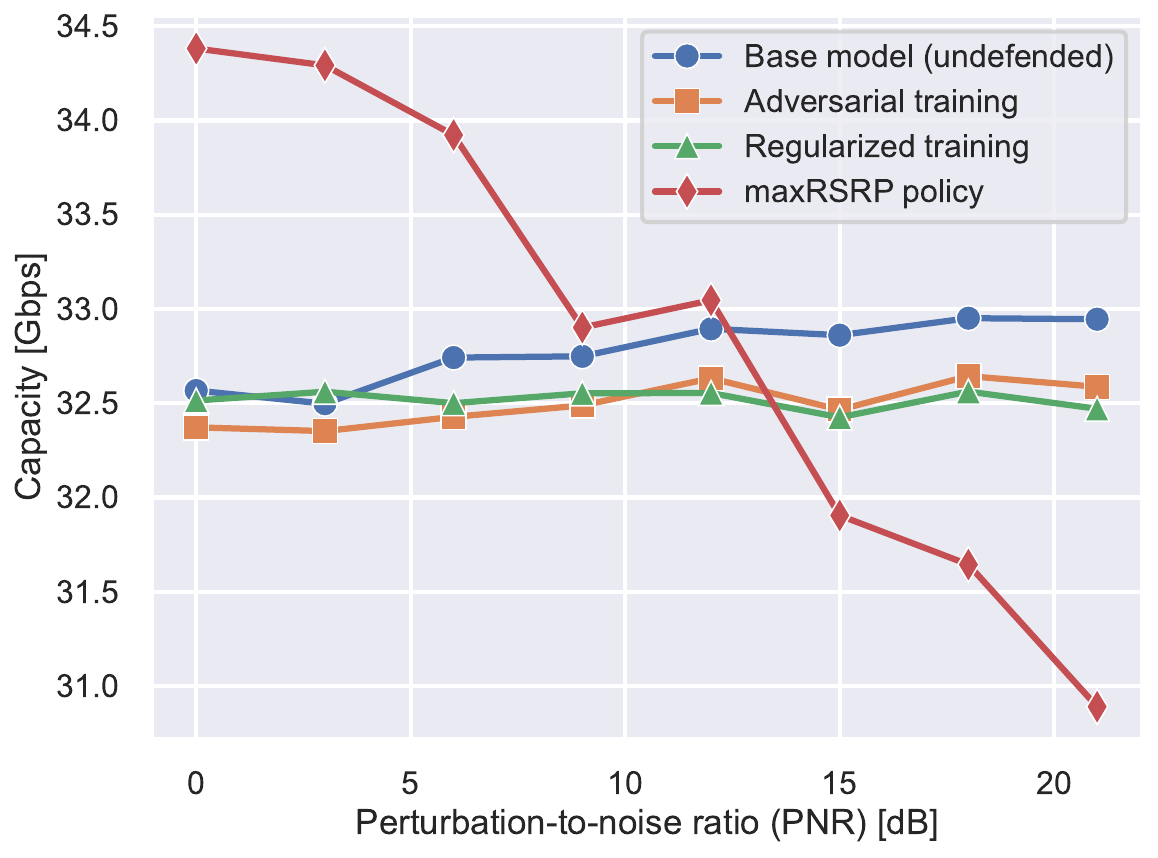}
    \caption{Capacity of maxRSRP algorithm drops with increase noise but the Connection management approach does not get impact. This is owing to the attack objective that primarily targets drop in the coverage performance.}
    \label{fig8}
\end{figure}

To observe the impact of noise to the maxRSRP policy, we add uniform random noise to the RSRP measurements with the same PNR budgets. We plot the coverage rate vs PNR for the maxRSRP policy as the average over eight different noise instances drawn from the uniform distribution. We observe that the maxRSRP policy encouters an increase in coverage rate (instead of decrease) with added noise and especially with high noise budget. This is owing to the fact that the policy becomes closer to a random allocation which is inherently more fair but can suffer from the overall capacity drop. This is illustrated in Figure~\ref{fig8}. It is interesting to note that while the maxRSRP policy encounters an increase in coverage rate with a corresponding decrease in the overall network capacity. However, the GNN-RL based connection management is unaffected in terms of the network capacity while only impacting the coverage, making network diagnostics challenging.

\begin{figure}[htbp]
    \centering
    \includegraphics[width=0.9\linewidth]{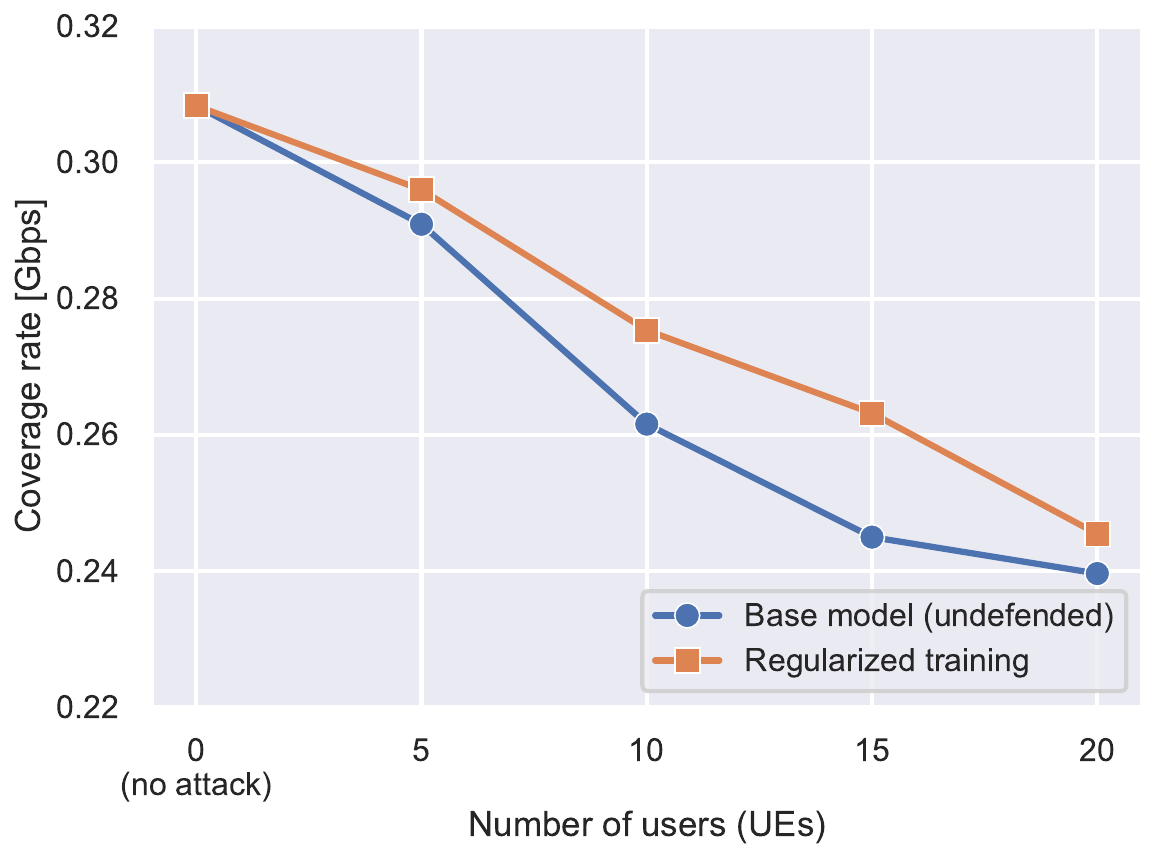}
    \caption{Whitebox Patch attack where the attacker is only allowed to perturb a subset of UEs' RSRP measurements. The horizontal axis shows the number of UEs' measurements that the attacker is allowed to perturb.}
    \label{fig7}
\end{figure}

We finally plot the performance of a patch attack where adversaries can only perturb a subset of UEs' RSRP measurements. To conduct this attack, we fix the perturbation budget to 20dB for each measurement, but limit the number of UEs that can be perturbed. Figure~\ref{fig7} plots the coverage performance as a function of the patch width or the number of UEs that are allowed to be perturbed by adversaries. The base model (undefended) faces a steep drop in coverage rate as the patch width increases to 40\% (20 UEs) of the total UEs in the network beyond which the rate of drop is smaller. For a patch width of 10 UEs (20\%) and 20 UEs, the base model suffers a 15\% and 22\% drop in the coverage rates respectively.

\subsection{Evaluating the Defenses}
We then conduct the same attacks but after training the models using the two defenses that were described in section~\ref{defenses}. For both adversarial training and regularized training approaches, we utilize PGD to obtain the worst-case perturbations. For the case of adversarial training, we simply utilize the set of rolled out experiences resulting from the perturbations into the replay buffer and update the DQN parameters. For the case of regularized training, we utilize the worst-case perturbations to define the policy regularizer term that is added to the original DQN loss objective at each training step. 

We revisit the coverage rate under physical attack in Figure~\ref{fig5a},~\ref{fig5b}, this time comparing the base model with 2 other trained models namely adversarially trained and the regularized models. We observe that both techniques result in significant recovery of performance of about 15\% in the coverage rate. Crucially, the regularized approach sustained such gains even under a high perturbation budget of $21dB$, while the adversarial training approach reduces to the case of the undefended model at very high noise regime.

When applying the two defense techniques, we assume that all entries of the RSRP matrix can be impacted. However, we also note that a robust model using regularized training in particular can also provide non-negligible gains in comparison to the base model under a patch attack as demonstrated in Figure~\ref{fig7}.

%For both adversarial training and regularized training, we did not conduct patch attacks and instead apply perturbation to the entire RSRP matrix while computing the noised states. Therefore, we observe that in the vanilla physical attack, both techniques result in significant recovery of performance of about 15\% in the coverage rate. Crucially, the regularized approach sustained such gains even under extreme noise conditions of PNR=21dB. It is interesting to note that when evaluating the impact of patch attacks in Figure 6 (right hand side), we observed that the regularized training also made the model more robust to patch attacks under different patch widths.

\section{Conclusions}
In this work, we utilized a practical AI based xApp for O-RAN system as an example to model the threat surfaces for adversaries to conduct evasion attacks that includes the wireless medium itself. We formulated new attack objectives for the attacker and successfully demonstrated the effectiveness of evasion attacks for a Connection Management xApp that utilized a Graph Neural Network trained using reinforcement learning. In particular, we showed the effectiveness of digital attacks (up to $50\%$ drop in coverage rate) and physical attacks (up to 25\% drop in coverage rate) using the above formulation. Finally, we designed robust training strategies using recent advances in adversarial machine learning for reinforcement learning and showed that the robustly trained models realize about 15\% improvement over undefended models for a range of perturbation budgets. Our findings can be broadly applicable to ML based xApps that utilize graph neural networks and reinforcement learning for optimizing wireless networks. In the future, we seek to rigorously analyze the robustness guarantees of different defenses and develop novel defenses that achieve strong robustness guarantees.

%\section*{Acknowledgment}

%\vspace{12pt}
\end{document}